\documentclass[conference]{IEEEtran}
\usepackage{setspace}
\usepackage{clrscode}
\usepackage{amsmath}
\usepackage{amssymb}
\usepackage[dvips]{graphicx}
\usepackage{epsfig}
\usepackage{hyperref}
\usepackage{bm}
\usepackage{subfigure}
\usepackage{verbatim}
\usepackage{titletoc}
\usepackage{extarrows}
\usepackage{fancyhdr}

\usepackage{balance}

\newcommand{\C}{{\sf{C}}}

\newcommand{\figsize}{0.45}

\newtheorem{Lem1}{Proposition}

\newtheorem{Def}{Definition}
\begin{document}

\title{Alternate Distributed Beamforming for Buffer-Aided Multi-Antenna Relay Systems}
    \author{
\IEEEauthorblockN{Jiayu Zhou\IEEEauthorrefmark{1}, Deli Qiao\IEEEauthorrefmark{1}, and Haifeng Qian\IEEEauthorrefmark{2}}
\IEEEauthorblockA{\IEEEauthorrefmark{1}\small{School of Communication and Electronic Engineering, East China Normal University, Shanghai, China}}
\IEEEauthorblockA{\IEEEauthorrefmark{2}\small{School of Computer Science and Software Engineering, East China Normal University, Shanghai, China}}
\small{Email: 52191214003@stu.ecnu.edu.cn, dlqiao@ce.ecnu.edu.cn, hfqian@cs.ecnu.edu.cn}}

\maketitle

\begin{abstract}\footnote{This work has been supported in part by the National Natural Science Foundation of China (61671205).}
In this paper, link selection is investigated in half-duplex (HD) dual-hop cooperative systems with multiple antennas at the relays. Alternate distributed beamforming (ADB) scheme is revisited for buffer-aided multi-antenna relay systems, in which the relays are divided into two groups, with one group receiving the same information broadcast from the source and the other group transmitting the common messages to the destination via distributed beamforming in each time slot. It is worth noting that the relays used for reception and transmission are determined without the need of instantaneous channel state information (CSI). Theoretical analysis of the achievable throughput of the proposed scheme in Rayleigh fading is provided and the approximate closed-form expressions are derived. Simulation results are given to verify the theoretical analysis.
Through numerical results, it is shown that compared with existing link selection policies, the fixed scheduling ADB scheme achieves a significant improvement in achievable throughput. It is also shown that for the ADB scheme, the throughput performance of increasing the number of antennas equipped at each relay is better than that of increasing the number of relays equipped with a single antenna when the total number of antennas at the relays is fixed.
\end{abstract}

\section{Introduction}
Nowadays, cooperative communications have attracted enormous attention in the research community, in which the communication between a source node and a destination node is assisted by one or more relay nodes
\cite{cooperative communication}. The diversity gain of the network can be obtained due to the alternative and independent transmission paths offered by the relays, and the distributed beamforming gain can also be expected \cite{DF}, \cite{ADB}, thus increasing the throughput and extending the coverage of cellular networks.

Various relay selection strategies have been proposed to better utilize the benefits provided by multiple relays. The conventional relay selection (CRS) scheme selects the relay with the strongest end-to-end signal to-noise ratio (SNR) \cite{CRS}. However, it fails to exploit the best source-relay channel and the best relay-destination channel at the same time on account of the fact that the relays are capable of storing data packets. The adoption of buffer-aided relays can provide both throughput and diversity gain by adaptive link selection \cite{buffer1}, \cite{buffer2}. Storing packets and transmitting them in favorable wireless conditions increases the network's resiliency, throughput and diversity. A max-max relay selection (MMRS) scheme was proposed in \cite{max-max}, where the relays for best reception and best transmission are selected respectively. In \cite{mimick}, a space full-duplex max-max relay selection (SFD-MMRS) scheme was introduced, which mimics full-duplex (FD) relaying with half-duplex (HD) relays via link selection. A new relay selection scheme called max-link selection scheme (MLS) was suggested in \cite{max-link}, which selects the strongest link for transmission among all the available links at each time slot. In addition, a general relay selection factor including the weight of the link and the link quality was proposed in \cite{weight}. It is worth noting that the relay selected with the adaptive link selection policies varies with the instantaneous channel state information (CSI).

Note that a relay usually operates in either full-duplex (FD) or half-duplex (HD) mode. In FD relaying, the relays transmit and receive at the same time and frequency, at the cost of hardware complexity \cite{FD1}, \cite{FD2}. We consider HD relays in this paper. However, the prelog factor $\frac{1}{2}$ is in the capacity expression due to the fact that relays are incapable of transmitting and receiving simultaneously, thus leading to reduced capacity of the whole network. Inspired by the decode-and-forward (DF) \cite{DF} and fixed scheduling \cite{1}, we have proposed an alternate distributed beamforming (ADB) scheme in \cite{ADB} for buffer-aided multi-relay systems to recover the HD loss with single antenna relays.

In this paper, we investigate the ADB scheme in a multi-antenna scenario. It is noted that due to the joint impacts of the multiple antennas and the number of relays, the analysis becomes much more challenging. We consider the HD buffer-aided cooperative multi-antenna relay systems. We assume that there is no direct link between the source and the destination. We analyze the achievable throughput of the ADB scheme in Rayleigh fading channels and derive the closed-form expressions, which offers an efficient way to measure the joint impacts of multiple antennas and the number of relays. Numerical results in accordance with theoretical analysis show the superiority of ADB in dual-hop cooperative systems with multiple antennas.

The reminder of this paper is organized as follows. The system model is introduced and several existing protocols are revisited in Section II. In Section III, the operation of ADB is briefly described, comprehensive analysis of the achievable throughput is presented and the approximate closed-form expressions are derived. Numerical results are provided in Section IV. Finally, conclusions are drawn in Section V with some lengthy proofs in Appendix.

\section{Preliminaries}
\begin{figure}
    \centering
    \includegraphics[width=0.4\textwidth]{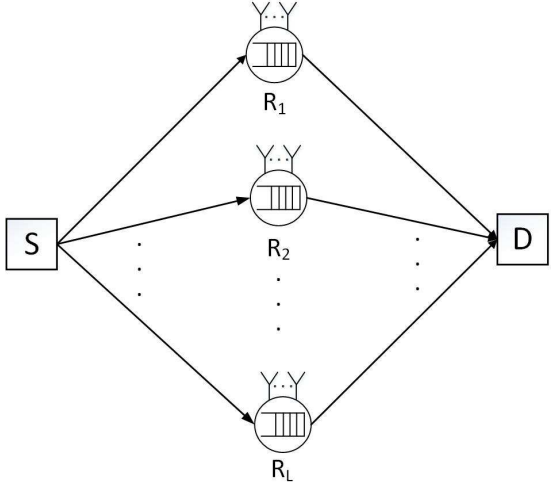}
    \caption{System model.}
    \label{fig:figure1}
\end{figure}
\subsection{System Model}
We consider a dual-hop cooperative multi-antenna relay systems with one source node $S$ and one destination node $D$ equipped with a single antenna, a set of $L$ DF relays $R_{1},...,R_{L}$ each with $N_{R}$ antennas, as shown in Fig. \ref{fig:figure1}. We assume that all nodes are operate in the HD mode, i.e., they cannot transmit and receive data simultaneously. We assume that there is a buffer of infinite length at each relay such that each relay can store the information received from the source and transmit it in later time.

Due to high path loss or shadowing effect, we assume that there is no direct link between the source and destination, and the communications can be achieved only through relays. Specifically, all the information that the destination receives is first processed by the relays. We use $h_{SR_{ij}}$ and $h_{R_{ij}D}$ for $i\in\{1,...,L\}$ and $j\in\{1,...,N_{R}\}$ to denote the channel coefficients of $S-R_{ij}$ and $R_{ij}-D$ links, respectively, where $R_{ij}$ denotes the $j$-th antenna of the relay $R_{i}$. The channel is assumed to be stationary and ergodic. We consider the block fading, in which the channel coefficients remain constant during one time slot and vary independently from one to the other.

We assume Rayleigh fading for the channel coefficients and the variances of $h_{SR_{ij}}$ and $h_{R_{ij}D}$ are assumed to be $\sigma_{h_{SR_{ij}}}^2$ and $\sigma_{h_{R_{ij}D}}^2$, respectively. Throughout this paper, we consider the case of independent and identically distributed (i.i.d.) fading for both $S-R_{ij}$ and the $R_{ij}-D$ links, i.e. $\sigma_{h_{SR_{ij}}}^2=\sigma_{g}^{2}$ and $\sigma_{h_{R_{ij}D}}^2=\sigma_{h}^{2}, i\in\{1,...,L\}, j\in\{1,...,N_{R}\}$ to facilitate the ensuing analysis \cite{mimick}.

\subsection{Two-Hop Transmission}
The transmission between $S$ and $D$ through relay $R_{i}$ is divided into two hops. In the first hop, relay $R_i$ receives data from $S$
and decodes with the maximal ratio combining (MRC) protocol. Hence, the received signals at the relay $R_i$ is given by
\begin{align}
y_{SR_{i}}= \textbf{h}_{SR_{i}}^{H}\textbf{w}_{SR_{i}}x_{S}+n_{R_{i}}.
\end{align}
where $x_{S}$ represents the signal transmitted by $S$ with an average power $P_{S}$. $\textbf{h}_{SR_{i}}=[h_{SR_{i,1}},h_{SR_{i,2}}\cdots h_{SR_{i,N_{R}}}]^{T}$ denotes the channel vectors between the $S$ and $R_{i}$, with $i\in\{1,...,L\}$. $\textbf{w}_{SR_{i}}=\textbf{h}_{SR_{i}}/\|\textbf{h}_{SR_{i}}\|$ is the receiving vector at $R_{i}$. $n_{R_{i}}$ is the additive white Gaussian noises (AWGNs) at $R_i$ with zero mean and variance $\sigma_{R}^{2}$. The instantaneous received signal-to-noise ratio (SNR) at relay $R_i$ is given by $\gamma_{SR_{i}}=\frac{P_{S}\|\textbf{h}_{SR_{i}}\|^2}{\sigma_{R}^{2}}$.
Note that $(\cdot)^{T}$ and $(\cdot)^{H}$ are denoted as the transpose and the conjugate transpose, and $\|\cdot\|$ is the Euclidean or $L_2$ vector norm.

Similarly, in the second hop, the relay $R_i$ transmits data to $D$ with the maximal ratio transmission (MRT) protocol. The received signals at $D$ is given as
\begin{align}
y_{R_{i}D}= \textbf{h}_{R_{i}D}^{H}\textbf{w}_{R_{i}D}x_{R}+n_{D}.
\end{align}
where $x_{R}$ is the signal transmitted by $R_i$ with an average power $P_{R}$. $\textbf{h}_{R_{i}D}=[h_{R_{i,1}D},h_{R_{i,2}D}, \cdots, h_{R_{i,N_{R}}D}]^{T}$ denotes the channel vectors between the $R_{i}$ and $D$, where $i\in\{1,...,L\}$. $\textbf{w}_{R_{i}D}=\textbf{h}_{R_{i}D}/\|\textbf{h}_{R_{i}D}\|$ is the transmit beamforming vectors at $R_{i}$. $n_{D}$ is the AWGNs at $D$ with zero mean and variance $\sigma_{D}^{2}$. And the instantaneous received SNR at the destination from relay $R_i$ is given by $\gamma_{R_{i}D}=\frac{P_{R}\|\textbf{h}_{R_{i}D}\|^2}{\sigma_{D}^{2}}$.
Without loss of generality, we assume that the noise power at the receiving nodes are equal to one, i.e., $\sigma_{R}^{2}=\sigma_{D}^{2}=1$.

\subsection{Existing Relaying Protocols}
In this part, we review several existing relaying protocols in the multi-antenna scenario. It is assumed that CSI is known at the transmitter of each link.

\subsubsection{Conventional Relay Selection (CRS)}
The conventional relay selection protocol selects the relay which provides the strongest end-to-end path between the source and destination \cite{CRS}. The source transmits in the first time slot and the selected relay forwards the data received from the source towards the destination in the second time slot. The best relay $R_j$ is selected based on
\begin{align}
j= {\rm arg} \max\limits_{i\in\{1,...,L\}}\{{\rm min}\{\gamma_{SR_{i}},\gamma_{R_{i}D}\}\}.
\end{align}
The instantaneous end-to-end capacity for the overall system is given by
\begin{align}
C_{k}=\frac{1}{2}{\rm log_{2}}\left(1+{\rm \max\limits_{1\leq \emph{k} \leq{L}}}{\rm min}\left(P_{S}\|\textbf{h}_{SR_{k}}\|^{2}, P_{R}\|\textbf{h}_{R_{k}D}\|^{2}\right)\right).
\end{align}
Then, the achievable throughput is given by ${\mathbb E}[C_{k}]$, where ${\mathbb E}[\cdot]$ denotes the expectation. Throughout this text, the unit for the throughput is bps/Hz.

\subsubsection{Space Full-Duplex Max-Max Relay Selection (SFD-MMRS)}
This protocol chooses different relays for reception and transmission, according to the quality of the channels, so that the relay selected for reception and the relay selected for transmission can receive and transmit at the same time \cite{mimick}. The best relay for reception $R_{r_{1}}$ and the best relay for transmission $R_{t_{1}}$ are selected respectively based on
\begin{align}
r_1={\rm arg} \max\limits_{i\in\{1,...,L\}}\{\gamma_{SR_{i}}\},
\end{align}
\begin{align}
t_1={\rm arg} \max\limits_{i\in\{1,...,L\}}\{\gamma_{R_{i}D}\}.
\end{align}
The second best relay for reception $R_{r_{2}}$ and the second best relay for transmission $R_{t_{2}}$ are selected respectively according to
\begin{align}
r_2=\arg \max_{\underset{i\neq r_1}{i\in \{1, \ldots, L\}}}  \{\gamma_{SR_{i}}\},
\end{align}
\begin{align}
t_2=\arg \max_{\underset{i\neq t_1}{i\in \{1, \ldots, L\}}}  \{\gamma_{R_{i}D}\}.
\end{align}
Then, in SFD-MMRS, the relays selected for reception $R_{\bar{r}_1}$ and transmission $R_{\bar{t}_1}$ are chosen as
\begin{align}
(R_{\bar{r}_1}, R_{\bar{t}_1})=
\begin{cases}
(R_{r_1}, R_{t_1}), \text{if} \quad r_1\neq t_1\\
(R_{r_2}, R_{t_1}), \text{if} \quad r_1= t_1 \text{and} \min(\gamma_{SR_{r_2}}, \gamma_{R_{t_1}D})\\
                         \hspace {1.8cm}   > \min(\gamma_{SR_{r_1}}, \gamma_{R_{t_2}D})\\
(R_{r_1}, R_{t_2}),  \text{otherwise}.
\end{cases}
\end{align}
Let $C_{SR}$ and $C_{RD}$ denote the instantaneous capacities of the $S-R$ and $R-D$ links, respectively, i.e.,
\begin{align}
C_{SR}&={\rm log_{2}}\left(1+P_{S}\|\textbf{h}_{SR_{\bar{r}_1}}\|^{2}\right),\ \nonumber\\
C_{RD}&={\rm log_{2}}\left(1+P_{R}\|\textbf{h}_{R_{\bar{t}_1}D}\|^{2}\right).
\end{align}
The achievable throughput is given by ${\rm min}\{{\mathbb E}[C_{SR}], {\mathbb E}[C_{RD}]\}$.

\subsubsection{Decode and Forward (DF)}
In DF \cite{DF}, each relay must decode the common message transmitted by the source node and beamform their transmissions to the destination, which is also performed in two time slots. Then, the instantaneous rate for the overall system is given by
\begin{small}
\begin{align}
&C_{k}=
\ \nonumber\\
&\frac{1}{2}{\rm log_{2}}\left(1+{\rm min}\left(P_{S}\min\limits_{1\leq{k}\leq{L}}\|\textbf{h}_{SR_{k}}\|^{2},P_{R}\left(\sum\limits_{k=1}^{L}\|\textbf{h}_{R_{k}D}\|\right)^{2}\right)\right).
\end{align}
\end{small}
The achievable throughput is given by ${\mathbb E}[C_{k}]$.

\section{Alternating Decode-and-Forward Protocol}

\subsection{The transmission policy}
\begin{figure}
    \centering
    \includegraphics[width=\figsize\textwidth]{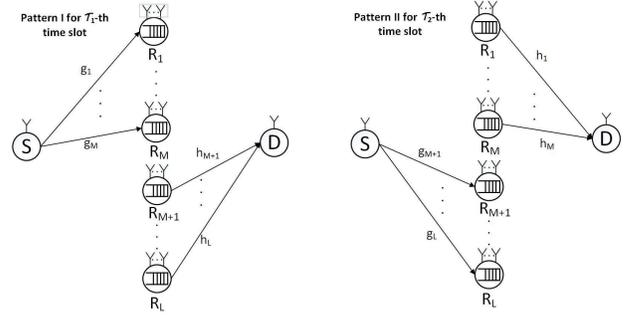}
    \caption{Transmission modes of the proposed scheme.}
    \label{fig:figure2}
\end{figure}
The operation of the ADB can be seen in Fig. \ref{fig:figure2}, which has two patterns. Time is slotted into discrete equal-size time slots. We divide $L$ relays into two groups, i.e., group 1 with $M$ relays, ${\mathcal R_{1}}=\{R_{1},...,R_{M}\}$ and group 2 with $L-M$ relays, ${\mathcal R_{2}}=\{R_{M+1},...,R_{L}\}$ \footnote{Due to the i.i.d. assumption, the relays can be divided arbitrarily. In case of different fading statistics, the relay grouping will be another interesting problem.}. The source broadcasts messages to the relays in group ${\mathcal R_{1}}$ for each ${t_{1}}$-th time slot while at the same time, the relays in group ${\mathcal R_{2}}$ beamform the data available in their buffers to the destination. It is assumed that the relays are synchronized through signaling. Similarly, during the ${t_{2}}$-th time slot, the relays in group ${\mathcal R_{2}}$ must decode the message transmitted by the source node and stores the packet in their buffers while the relays in group ${\mathcal R_{1}}$ beamform the previously received packets to the destination.

In this strategy, the benefits of both DF and fixed scheduling are enjoyed. It is obvious that with this protocol, the HD loss of conventional relays can be recovered \cite{mimick} and distributed beamforming gain can be expected. And it is worth noting that compared with the selective protocols CRS and SFD-MMRS, the receiving and transmitting relays in the proposed policy do not vary with the instantaneous CSI and are predetermined at the beginning of transmissions.

\subsection{Achievable Throughput Analysis}
In this section, we analyze the achievable throughput performance of the ADB scheme for buffer-aided multi-antenna relay systems and derive the approximate closed-form expressions. Due to the assumption of no inter-relay links \cite{mimick}, \cite{bypassing}, we assume that there is no inter-relay interference when the receiving relays and transmitting relays are active in the same time-slot. In practice, this assumption is valid if the relays are located far away from each other or if fixed infrastructure-based relays with directional antennas are used. Note that fixed relays are of practical interest since they are low-cost and low-transmit power devices (see, e.g., \cite{bypassing}, \cite{frelay2}, and \cite{frelay3}).
First, we have the following results.

Given the transmit power levels $P_{S}$ and $P_{R}$, the achievable throughput of the proposed scheme can be expressed as \cite{ADB}
\begin{small}
\begin{align}
C_{ADB}(P_{S},P_{R})
&=\frac{1}{2}{\rm min}\Bigg\{{\mathbb E}\left[{\rm log_{2}}\left(1+P_{S}\min\limits_{R_{i}\in{\mathcal R_{1}}}(\|\textbf{h}_{SR_{i}}\|^2)\right)\right],
\ \nonumber\\
&{\mathbb E}\left[{\rm log_{2}}\left(1+P_{R}\left(\sum\limits_{R_{i}\in{\mathcal R_{1}}}\|\textbf{h}_{R_{i}D}\|\right)^{2}\right)\right]\Bigg\}
\ \nonumber\\
&+\frac{1}{2}{\rm min}\Bigg\{{\mathbb E}\left[{\rm log_{2}}\left(1+P_{S}\min\limits_{R_{i}\in{\mathcal R_{2}}}(\|\textbf{h}_{SR_{i}}\|^2)\right)\right],
\notag
\end{align}
\end{small}
\begin{small}
\begin{align}
&{\mathbb E}\left[{\rm log_{2}}\left(1+P_{R}\left(\sum\limits_{R_{i}\in{\mathcal R_{2}}}\|\textbf{h}_{R_{i}D}\|\right)^{2}\right)\right]\Bigg\}.\label{ma-throughput2}
\end{align}
\end{small}
Same as \cite{ADB}, without loss of generality, we consider the case that two different transmission modes alternates every time slot.

Denote
\begin{align}
&C_{11}={\mathbb E}\left[{\rm log_{2}}\left(1+P_{S}\min\limits_{R_{i}\in{\mathcal R_{1}}}(\|\textbf{h}_{SR_{i}}\|^2)\right)\right],\ \\
&C_{12}={\mathbb E}\left[{\rm log_{2}}\left(1+P_{R}\left(\sum\limits_{R_{i}\in{\mathcal R_{2}}}\|\textbf{h}_{R_{i}D}\|\right)^{2}\right)\right],\ \\
&C_{21}={\mathbb E}\left[{\rm log_{2}}\left(1+P_{S}\min\limits_{R_{i}\in{\mathcal R_{2}}}(\|\textbf{h}_{SR_{i}}\|^2)\right)\right],\ \\
&C_{22}={\mathbb E}\left[{\rm log_{2}}\left(1+P_{R}\left(\sum\limits_{R_{i}\in{\mathcal R_{1}}}\|\textbf{h}_{R_{i}D}\|\right)^{2}\right)\right].
\end{align}

\begin{Lem1}\label{prop:ma-closed-form}
Given $P_{S}$ and $P_{R}$, the approximate closed-form expressions for the achievable throughput of ADB in Rayleigh fading channels are given by
\end{Lem1}
\begin{align}
\C_{ADB}&=\frac{1}{2}{\rm min}(\C_{11},\C_{22})+\frac{1}{2}{\rm min}(\C_{21},\C_{12}) \nonumber\\
&\hspace{-.5cm}=\left\{\begin{array}{ll}
\frac{1}{2}(\C_{11}+\C_{21}) & \text{if}\quad \C_{11}<\C_{22}\quad \text{and} \quad \C_{21}<\C_{12}, \\
\frac{1}{2}(\C_{11}+\C_{12}) & \text{if}\quad \C_{11}<\C_{22}\quad \text{and} \quad \C_{21}>\C_{12}, \\
\frac{1}{2}(\C_{22}+\C_{21}) & \text{if}\quad \C_{11}>\C_{22}\quad \text{and} \quad \C_{21}<\C_{12}, \\
\frac{1}{2}(\C_{22}+\C_{12}) & \text{otherwise}.\label{ma-4cases}
\end{array}\right.
\end{align}
where
\begin{small}
\begin{align}
&\C_{11}=\sum_{\substack{n_{i}\geq0,\\ n_{1}+n_{2}+...+n_{N_{R}}=M}}\frac{\binom{M}{n_{1},n_{2},...,n_{N_{R}}} \cdot\left(\frac{1}{2\sigma^{2}_{g}}\right)^{p}}{qln2}\bigg[\left(-\frac{1}{P_{S}}\right)^{p}
\ \nonumber\\
&e^{\frac{M}{2P_{S}\sigma_{g}^{2}}}E_{1}\left(\frac{M}{2P_{S}\sigma_{g}^{2}}\right)+\sum\limits_{r=1}^{p}(r-1)!\left(-\frac{1}{P_{S}}\right)^{p-r}\left(\frac{M}{2\sigma_{g}^{2}}\right)^{-r}\bigg],\ \\
&\C_{22}=\sum\limits_{k=0}^{N_{R}M-1}\frac{1}{k!\left(\frac{1}{2M\sigma_{h}^{2}}\right)^{-k}\ln2}\bigg[\left(-\frac{1}{P_{R}}\right)^{k}e^{\frac{1}{2P_{R}M\sigma_{h}^{2}}}
\ \nonumber\\
&E_{1}(2P_{R}M\sigma_{h}^{2})+\sum\limits_{r=1}^{k}(r-1)!\left(-\frac{1}{P_{R}}\right)^{k-r}\left(\frac{1}{2M\sigma_{h}^{2}}\right)^{-r}\bigg],\ \\
&\C_{21}=\sum_{\substack{n_{i}\geq0,\\ n_{1}+n_{2}+...+n_{N_{R}}=L-M}}\frac{\binom{L-M}{n_{1},n_{2},...,n_{N_{R}}} \cdot(\frac{1}{2\sigma^{2}_{g}})^{p}}{qln2}\bigg[\left(-\frac{1}{P_{S}}\right)^{p}
\ \nonumber\\
&e^{\frac{L-M}{2P_{S}\sigma_{g}^{2}}}E_{1}\left(\frac{L-M}{2P_{S}\sigma_{g}^{2}}\right)
+\sum\limits_{r=1}^{p}(r-1)!\left(-\frac{1}{P_{S}}\right)^{p-r}\left(\frac{L-M}{2\sigma_{g}^{2}}\right)^{-r}\bigg],\ \\
&\C_{12}=\sum\limits_{k=0}^{N_{R}(L-M)-1}\frac{1}{k!\left(\frac{1}{2(L-M)\sigma_{h}^{2}}\right)^{-k}\ln2}
\notag
\end{align}
\end{small}
\begin{small}
\begin{align}
&\bigg[\left(-\frac{1}{P_{R}}\right)^{k}e^{\frac{1}{2P_{R}(L-M)\sigma_{h}^{2}}}E_{1}\left(2P_{R}(L-M)\sigma_{h}^{2}\right)
\ \nonumber\\
&+\sum\limits_{r=1}^{k}(r-1)!\left(-\frac{1}{P_{R}}\right)^{k-r}\left(\frac{1}{2(L-M)\sigma_{h}^{2}}\right)^{-r}\bigg],
\end{align}
\end{small}
with $p=\sum\limits_{i=0}^{N_{R}-1}i\cdot n_{i+1}$, $q=\prod\limits_{i=0}^{N_{R}-1}(i!)^{n_{i+1}}$,  $\binom{M}{n_{1},n_{2},...,n_{N_{R}}}=\frac{M!}{n_{1}!n_{2}!\cdots n_{N_{R}}!}$, and $E_{1}(x)=\int_{x}^\infty(e^{-t}/t)dt, x>0$ is the exponential integral function.

\emph{Proof:} Please see Appendix \ref{app:ma-closed-form}.\hfill$\square$

Given the total power constraint SNR of the network, we can allocate the total power to the source and relays to achieve the best performance.

For ADB, the source transmits in every time slot, while either $M$ relays in group ${\mathcal R_{1}}$ or $L-M$ relays in group ${\mathcal R_{2}}$ transmits in one time slot alternatively. Therefore, we should have $P_{S}+\frac{L}{2}P_{R} \leq {\rm SNR}$. Regarding SFD-MMRS, we allocate transmit power to the source and $L$ relays to enable each relay to be capable of being selected for transmission. Again, the sources works for all time slots. So we should have $P_{S}+LP_{R} \leq {\rm SNR}$. With CRS, similarly, we should allocate transmit power to the source and $L$ relays, albeit the data transmission occupies two time slots. Therefore, we should have $\frac{1}{2}(P_{S}+LP_{R}) \leq {\rm SNR}$. With regard to DF, each relay must decode the common message transmitted by the source node and beamform their transmissions to the destination, obviously we need to allocate transmit power to source and $L$ relays. It is also performed in two time slots, so we should have $\frac{1}{2}(P_{S}+LP_{R}) \leq {\rm SNR}$.

Consider the achievable throughput in (\ref{ma-throughput2}), once given the total power SNR, it is obvious that when $P_{S}$ is small, the throughput is limited by the source-relay link. On the other hand, when $P_R$ is small, the relay-destination link will be the bottleneck of the system. Therefore, there is always an optimal power allocation that maximizes the achievable throughput.
\begin{Def}
The maximum achievable throughput of ADB is given by
\begin{align}
C_{max}=\max\limits_{P_{S}+\frac{L}{2}P_{R} \leq SNR}C_{ADB}(P_{S},P_{R}).
\end{align}
\end{Def}
Similarly, we can define the maximum achievable throughput for DF, CRS, and SFD-MMRS.

\section{Numerical Results}

In this section, we evaluate the proposed ADB scheme and compare it with that of CRS \cite{CRS}, SFD-MMRS \cite{mimick}, and DF \cite{DF}. We assume that $\sigma_{g}^{2}=\sigma_{h}^{2}=1$, unless specified otherwise.

\begin{figure}
    \centering
    \includegraphics[width=\figsize\textwidth]{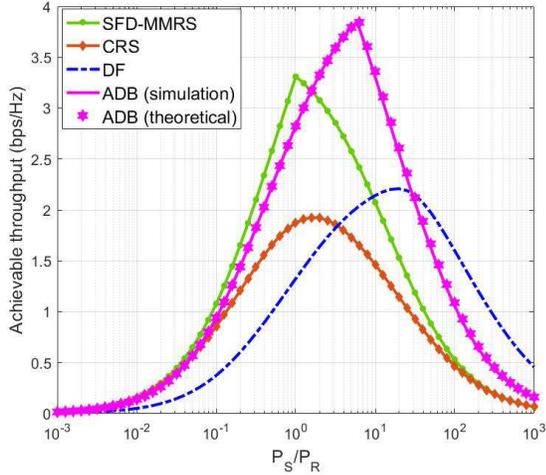}
    \caption{Achievable throughput versus $P_{S}/P_{R}$ for several relaying protocols.}
    \label{figure3}
\end{figure}
Fig. \ref{figure3} plots the achievable throughput versus $P_{S}/P_{R}$ for each scheme. We assume ${\rm SNR}=10$ dB, $N_{R}=3$, $L=4$ and $m=2$. We can find that the achievable throughput always has a peak value as $P_{S}/P_{R}$ varies,
which verifies that once given the total power SNR, there is always an optimal power allocation that maximizes the achievable throughput, and the achievable throughput corresponding to the optimal power allocation, rather than the throughput corresponding to each power ratio, is what we are concerned about. We can see that the proposed scheme achieves the largest throughput. We also note that the analytical results obtained based on the derivation in Section III match the simulation results, which verifies the approximate closed-form expressions.

\begin{figure}
    \centering
    \includegraphics[width=\figsize\textwidth]{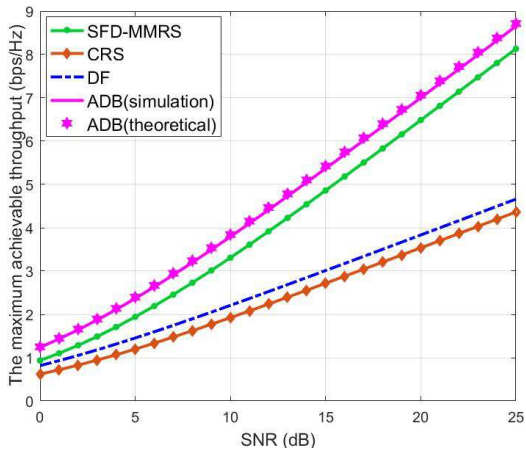}
    \caption{The maximum achievable throughput versus SNR for several relaying protocols.}
    \label{figure4}
\end{figure}
In Fig. \ref{figure4}, we compare the maximum achievable throughput of the proposed ADB scheme with that of two relay selection schemes and the traditional DF scheme as SNR varies. We assume $N_{R}=3$, $L=4$ and $m=2$. We can find that the proposed scheme achieves the best performance in all cases. Also, we can find that the approximate expression holds for a wide range of SNR values.

\begin{figure}
    \centering
    \includegraphics[width=\figsize\textwidth]{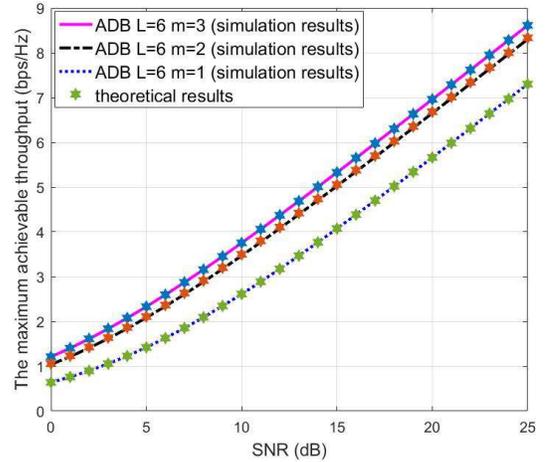}
    \caption{The maximum achievable throughput versus SNR for different grouping modes of the proposed scheme.}
    \label{figure5}
\end{figure}
In Fig. \ref{figure5}, we compare the maximum achievable throughput of the ADB scheme versus SNR for different $m$, i.e., different grouping modes. We assume $N_{R}=3$ and $L=6$. It is interesting that, the symmetric allocation of relays achieves the best performance with the given setting. This is generally because that beamforming gain can be attained within each group.

\begin{figure}
    \centering
    \includegraphics[width=\figsize\textwidth]{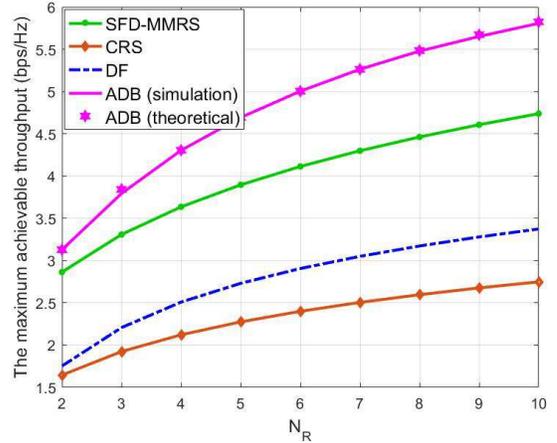}
    \caption{The maximum achievable throughput versus the number of antennas equipped at each relay for several relaying protocols.}
    \label{figure6}
\end{figure}
In Fig. \ref{figure6}, we plot the maximum achievable throughput of each scheme versus the number of antennas of each relay for ${\rm SNR}=10$ dB. We assume $L=4$ and $m=2$. We can clearly see that the maximum achievable throughput improves as $N_{R}$ increases. And the proposed scheme achieves a significant improvement in achievable throughput. We also observe that as $N_{R}$ increases, the superiority of the proposed scheme over other strategies in achievable throughput becomes more apparent.

\begin{figure}
    \centering
    \includegraphics[width=\figsize\textwidth]{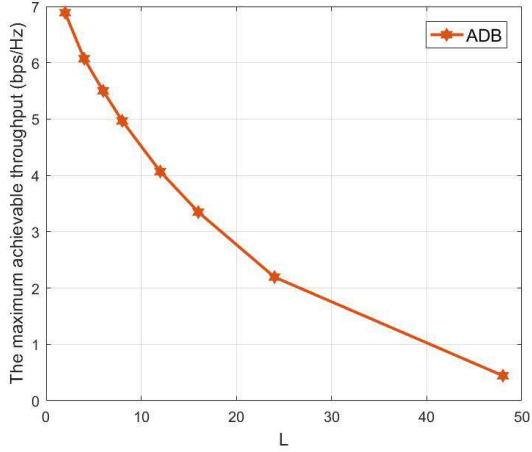}
    \caption{The maximum achievable throughput versus the number of relays for several relaying protocols.}
    \label{figure7}
\end{figure}
Fig. \ref{figure7} plots the maximum achievable throughput versus the number of relays $L$ for a fixed total number of antennas. We assume $m=\frac{L}{2}$, and the total number of antennas is fixed as $N_{t}=48$. It is interesting that the maximum achievable throughput decreases as $L$ increases, i.e., compared with only single antenna for the relay, the more antennas the better for the given total number of antennas. In other words, when the number of antennas is constrained, it is better to increase the number of antennas per relay rather than the number of single antenna relays.

\section{Conclusion}
In this paper, we have investigated ADB for buffer-aided cooperative multi-antenna relaying systems, in which the relays are divided into two predetermined and unaltered groups, with one group receiving the signals transmitted by the source node while at the same time, the other group beamforming the previously received data to the destination. We have obtained the closed-form expressions of the achievable throughput in Rayleigh fading channels. Through numerical results, we have found that the ADB scheme achieves significant improvement over the existing schemes in terms of the maximum achievable throughput. In addition, for a given total number of antennas, we have found that the proposed scheme achieves larger
throughput with the increasing number of antennas per relay rather than the number of single antenna relays.

\appendix
\subsection{Proof of Proposition \ref{prop:ma-closed-form}}\label{app:ma-closed-form}
To compute the achievable throughput of ADB in (\ref{ma-4cases}), we need to find $\C_{11}$, $\C_{12}$, $\C_{21}$, and $\C_{22}$.

\quad $Computation$ $of$ $\C_{11}$: In this case, we denote $z=\min\limits_{i\in\{1,...,M\}}t_{i}$, where $t_{i}=\|\textbf{h}_{SR_{i}}\|^2$.
Therefore, to derive $\C_{11}$, we first compute the cumulative distribution function (CDF) of $z$.
Denote $z={\rm min}(\|\textbf{h}_{SR_{1}}\|^2,\|\textbf{h}_{SR_{2}}\|^2,...,\|\textbf{h}_{SR_{M}}\|^2)$.
$t_{i}$ follows the Erlang distribution. The CDF of $t_{i}$ is given by \cite[17.2]{distributions handbook}
\begin{align}
F_{T_i}(t_{i})=1-\sum\limits_{r=0}^{N_{R}-1}\frac{\left(\frac{t_{i}}{2\sigma_{g}^{2}}\right)^{r}e^{-\frac{t_{i}}{2\sigma_{g}^{2}}}}{r!}.
\end{align}
Then the CDF of $z$ can be calculated as
\begin{align}
F_{Z}(z)
&=P\left(\min\limits_{i\in\{1,...,M\}}t_{i}\leq z\right)=1-P\left(\min\limits_{i\in\{1,...,M\}}t_{i}\geq z\right)\ \nonumber\\
&=1-P(t_{1}\geq z)P(t_{2}\geq z)...P(t_{M}\geq z)\ \nonumber\\
&=1-\left(1-F_{T_i}(z)\right)^{M}=1-\left(\sum\limits_{r=0}^{N_{R}-1}\frac{\left(\frac{z}{2\sigma_{g}^{2}}\right)^{r}e^{-\frac{z}{2\sigma_{g}^{2}}}}{r!}\right)^{M}.\label{ma-zcdf}
\end{align}
Then $\C_{11}$ can be obtained as
\begin{small}
\begin{align}
\C_{11}
&={\mathbb E}\left[{\rm log_{2}}\left(1+P_{S}\min\limits_{i\in\{1,...,M\}}\|\textbf{h}_{SR_{i}}\|^2\right)\right]\ \nonumber\\
&={\mathbb E}[{\rm log_{2}}(1+P_{S}z)]=\int_{0}^{\infty}{\rm log_{2}}(1+P_{S}z)d\left(F_{Z}(z)-1\right)\ \nonumber\\
&=-{\rm log_{2}}(1+P_{S}z)\left(\sum\limits_{r=0}^{N_{R}-1}\frac{\left(\frac{z}{2\sigma_{g}^{2}}\right)^{r}e^{-\frac{z}{2\sigma_{g}^{2}}}}{r!}\right)^{M}\Bigg|_{0}^{\infty}
\ \nonumber\\
&+\int_{0}^{\infty}\left(\sum\limits_{r=0}^{N_{R}-1}\frac{\left(\frac{z}{2\sigma_{g}^{2}}\right)^{r}e^{-\frac{z}{2\sigma_{g}^{2}}}}{r!}\right)^{M}d(\log_{2}(1+P_{S}z))
\ \nonumber\\
&=\frac{1}{\rm ln2}\int_{0}^{\infty}\left(z+\frac{1}{P_{S}}\right)^{-1} e^{-\frac{M}{2\sigma_{g}^{2}}z}\left(\sum\limits_{r=0}^{N_{R}-1}\frac{\left(\frac{z}{2\sigma_{g}^{2}}\right)^{r}}{r!}\right)^{M}dz
\ \nonumber\\
&\overset{a}{=}\sum_{\substack{n_{i}\geq0,\\ n_{1}+n_{2}+...+n_{N_{R}}=M}}\frac{\binom{M}{n_{1},n_{2},...,n_{N_{R}}}\cdot \left(\frac{1}{2\sigma_{g}^{2}}\right)^{\sum\limits_{i=0}^{N_{R}-1}i\cdot n_{i+1}}}{\ln2\prod\limits_{i=0}^{N_{R}-1}(i!)^{n_{i+1}}}
\ \nonumber\\
&\int_{0}^{\infty}\frac{z^{\sum\limits_{i=0}^{N_{R}-1}i\cdot n_{i+1}}e^{-\frac{M}{2\sigma_{g}^{2}}z}}{z+\frac{1}{P_{S}}}dz
\ \nonumber\\
&\overset{b}{=}\sum_{\substack{n_{i}\geq0,\\ n_{1}+n_{2}+...+n_{N_{R}}=M}}\frac{\binom{M}{n_{1},n_{2},...,n_{N_{R}}}\cdot \left(\frac{1}{2\sigma_{g}^{2}}\right)^{\sum\limits_{i=0}^{N_{R}-1}i\cdot n_{i+1}}}{\ln2\prod\limits_{i=0}^{N_{R}-1}(i!)^{n_{i+1}}}
\ \nonumber\\
&\bigg[\left(-\frac{1}{P_{S}}\right)^{\sum\limits_{i=0}^{N_{R}-1}i\cdot n_{i+1}}e^{\frac{M}{2P_{S}\sigma_{g}^{2}}}E_{1}\left(\frac{M}{2P_{S}\sigma_{g}^{2}}\right)
\ \nonumber\\
&+\sum\limits_{r=1}^{\sum\limits_{i=0}^{N_{R}-1}i\cdot n_{i+1}}(r-1)!\left(-\frac{1}{P_{S}}\right)^{{\sum\limits_{i=0}^{N_{R}-1}i\cdot n_{i+1}}-r}\left(\frac{M}{2\sigma_{g}^{2}}\right)^{-r}\bigg],\label{ma-C11}
\end{align}
\end{small}
where polynomial theorem is used in equality (a) and Eq. 3.353.5 in \cite{integral} is used to obtain equality (b). And $E_{1}(x)=\int_{x}^\infty(e^{-t}/t)dt, x>0$ is the exponential integral function.

The computation of $\C_{21}$ is similar to that of $\C_{11}$, it is given by
\begin{small}
\begin{align}
\C_{21}
&=\sum_{\substack{n_{i}\geq0,\\ n_{1}+n_{2}+...+n_{N_{R}}=L-M}}\frac{\binom{L-M}{n_{1},n_{2},...,n_{N_{R}}}\cdot \left(\frac{1}{2\sigma_{g}^{2}}\right)^{\sum\limits_{i=0}^{N_{R}-1}i\cdot n_{i+1}}}{\ln2\prod\limits_{i=0}^{N_{R}-1}(i!)^{n_{i+1}}}
\notag
\end{align}
\end{small}
\begin{small}
\begin{align}
&\bigg[\left(-\frac{1}{P_{S}}\right)^{\sum\limits_{i=0}^{N_{R}-1}i\cdot n_{i+1}}e^{\frac{L-M}{2P_{S}\sigma_{g}^{2}}}E_{1}\left(\frac{L-M}{2P_{S}\sigma_{g}^{2}}\right)
\ \nonumber\\
&+\sum\limits_{r=1}^{\sum\limits_{i=0}^{N_{R}-1}i\cdot n_{i+1}}(r-1)!\left(-\frac{1}{P_{S}}\right)^{{\sum\limits_{i=0}^{N_{R}-1}i\cdot n_{i+1}}-r}\left(\frac{L-M}{2\sigma_{g}^{2}}\right)^{-r}\bigg].\label{ma-C21}
\end{align}
\end{small}
\quad $Computation$ $of$ $\C_{22}$: In this case, let $z=\sum\limits_{i=1}^{M}s_{i}$, where $s_{i}=\|\textbf{h}_{R_{i}D}\|, i\in\{1,...,M\}$.
Denote $t_{i}=\|\textbf{h}_{R_{i}D}\|^{2}, i\in\{1,...,M\}$, $t_{i}$ follows the Erlang distribution, the probability density function (PDF) of which is given by
\begin{align}
f_{T_{i}}(t_{i})=\frac{\left(\frac{1}{2\sigma_{h}^{2}}\right)^{N_{R}}t_{i}^{N_{R}-1}e^{-\frac{t_{i}}{2\sigma_{h}^{2}}}}{(N_{R}-1)!}.
\end{align}
The gamma distribution is a continuous probability distribution. When the shape parameter has an integer value, the distribution is the Erlang distribution. And it is well known that a Nakagami-$m$ random variable (RV) is the square root of a gamma RV, so we can know that $s_{i}$ follows the Nakagami distribution, the PDF of $s_{i}$ is correspondingly given by
\begin{align}
f_{S_{i}}(s_{i})=\frac{2\left(\frac{1}{2\sigma_{h}^{2}}\right)^{N_{R}}s_{i}^{2N_{R}-1}e^{-\frac{s_{i}^{2}}{2\sigma_{h}^{2}}}}{(N_{R}-1)!}.
\end{align}
We denote $s_{i}\sim\mathcal{M}(s, m, \Omega)$, where $m=N_{R}$ and $\Omega=2N_{R}\sigma_{h}^{2}$.
So far we know that $z$ is a sum of $M$ i.i.d. Nakagami random variables (RV's). A relatively simple and widely used approximation for the sum PDF was given in \cite[(82),(84)]{m-distribution}, from where we can have $z\simeq\mathcal{M}(z, {}_{0}m, {}_{0}\Omega)$, where ${}_{0}m\simeq mM$ and ${}_{0}\Omega\simeq M_{2}\Omega$.
Then the approximation for the PDF of $z$ is given by
\begin{align}
f_{Z}(z)=\frac{2\left(\frac{1}{2M\sigma_{h}^{2}}\right)^{N_{R}M}z^{2N_{R}M-1}e^{-\frac{z^{2}}{2M\sigma_{h}^{2}}}}{(N_{R}M-1)!},\label{ma-fSAA}
\end{align}
The CDF of $z$ is simply obtained by integrating the PDF in (\ref{ma-fSAA}) with respect to $z$ and is given by
\begin{align}
F_{Z}(z)
&=\int_{0}^{z}\frac{2\left(\frac{1}{2M\sigma_{h}^{2}}\right)^{N_{R}M}t^{2N_{R}M-1}e^{-\frac{t^{2}}{2M\sigma_{h}^{2}}}}{(N_{R}M-1)!}dt
\ \nonumber\\
&\xlongequal{x=t^{2}}\frac{\left(\frac{1}{2M\sigma_{h}^{2}}\right)^{N_{R}M}}{(N_{R}M-1)!}\int_{0}^{z^2}x^{N_{R}M-1}e^{-\frac{x}{2M\sigma_{h}^{2}}}dx
\ \nonumber\\
&=1-e^{-\frac{z^{2}}{2M\sigma_{h}^{2}}}\sum\limits_{k=0}^{N_{R}M-1}\frac{z^{2k}}{k!\left(\frac{1}{2M\sigma_{h}^{2}}\right)^{-k}}.
\end{align}
\allowdisplaybreaks[2]
Then the $\C_{22}$ can be computed as
\begin{small}
\begin{align}
\C_{22}
&={\mathbb E}\left[{\rm log_{2}}\left(1+P_{R}(\sum\limits_{i=1}^{M}\|\textbf{h}_{R_{i}D}\|)^{2}\right)\right]\ \nonumber\\
&={\mathbb E}\left[{\rm log_{2}}\left(1+P_{R}z^{2}\right)\right]=\int_{0}^{\infty}{\rm log_{2}}(1+P_{R}z^{2})f_{Z}(z)dz\ \nonumber\\
&={\rm log_{2}}(1+P_{R}z^{2})d(F_{Z}(z)-1)\ \nonumber\\
&=-\log_{2}(1+P_{R}z^{2})e^{-\frac{z^{2}}{2M\sigma_{h}^{2}}}\sum\limits_{k=0}^{N_{R}M-1}\frac{z^{2k}}{k!\left(\frac{1}{2M\sigma_{h}^{2}}\right)^{-k}}\bigg|_{0}^{\infty}
\ \nonumber\\
&+\int_{0}^{\infty}e^{-\frac{z^{2}}{2M\sigma_{h}^{2}}}\sum\limits_{k=0}^{N_{R}M-1}\frac{z^{2k}}{k!\left(\frac{1}{2M\sigma_{h}^{2}}\right)^{-k}}d\left({\rm log_{2}}(1+P_{R}z^{2})\right)\ \nonumber\\
&\xlongequal{x=z^{2}}\sum\limits_{k=0}^{N_{R}M-1}\frac{1}{k!\left(\frac{1}{2M\sigma_{h}^{2}}\right)^{-k}\ln2}\int_{0}^{\infty}\frac{x^{k}e^{-\frac{1}{2M\sigma^{2}_{h}}x}}{x+\frac{1}{P_{R}}}dx
\ \nonumber\\
&=\sum\limits_{k=0}^{N_{R}M-1}\frac{1}{k!\left(\frac{1}{2M\sigma_{h}^{2}}\right)^{-k}\ln2}\bigg[\left(-\frac{1}{P_{R}}\right)^{k}e^{\frac{1}{2P_{R}M\sigma_{h}^{2}}}E_{1}\left(\frac{1}{2P_{R}M\sigma_{h}^{2}}\right)
\ \nonumber\\
&+\sum\limits_{r=1}^{k}(r-1)!\left(-\frac{1}{P_{R}}\right)^{k-r}\left(\frac{1}{2M\sigma_{h}^{2}}\right)^{-r}\bigg],\label{ma-C22}
\end{align}
\end{small}
where Eq. 3.353.5 in \cite{integral} is used to obtain the final equality.

The computation of $\C_{12}$ is similar to that of $\C_{22}$, it is given by
\begin{small}
\begin{align}
\C_{12}
&=\sum\limits_{k=0}^{N_{R}(L-M)-1}\frac{1}{k!\left(\frac{1}{2(L-M)\sigma_{h}^{2}}\right)^{-k}\ln2}
\ \nonumber\\
&\Big[\left(-\frac{1}{P_{R}}\right)^{k}e^{\frac{1}{2P_{R}(L-M)\sigma_{h}^{2}}}E_{1}\left(\frac{1}{2P_{R}(L-M)\sigma_{h}^{2}}\right)
\ \nonumber\\
&+\sum\limits_{r=1}^{k}(r-1)!\left(-\frac{1}{P_{R}}\right)^{k-r}\left(\frac{1}{2(L-M)\sigma_{h}^{2}}\right)^{-r}\Big].\label{ma-C12}
\end{align}
\end{small}
Finally, $\C_{ADB}$ is obtained by substituting (\ref{ma-C11}), (\ref{ma-C21}), (\ref{ma-C22}), and (\ref{ma-C12}) into (\ref{ma-4cases}).

\balance

\end{document}